\documentstyle[prl,aps]{revtex}

\begin{document}

\twocolumn[{

\title{Huge Longitudinal Resistance in the Fractional Quantum Hall Effect Regime}
\author{S. Kronm\"uller,  W. Dietsche, J. Weis and K. v. Klitzing}
\address{Max-Planck-Institut f\"ur Festk\"orperforschung, Stuttgart, Germany}
\author{W. Wegscheider and M. Bichler}
\address{Walter Schottky Institut, Technische Universit\"at M\"unchen, M\"unchen, Germany }

\date{\today}
\maketitle

\begin{abstract}
\widetext
\leftskip 54.8 pt
\rightskip 54.8 pt
The magneto resistance of a narrow single quantum well is spectacularly different from
the usual behavior. At filling factors  $\rm \frac{2}{3}$ and $\rm \frac{3}{5}$  we observe 
large and sharp maxima in the longitudinal resistance instead of the expected minima. 
The peak value of the resistance exceeds those of the surrounding
magnetic field regions by a factor of up to three. 
The formation of the maxima takes place on very large time scales which suggests a close
relation with nuclear spins.
We discuss the properties of the observed maxima due to a formation of domains of different electronic states.

\end{abstract}
}]

\narrowtext
The integer and  fractional quantum Hall effects (IQHE \cite{1} and FQHE \cite{2})
 have been the subject of 
intense investigation for more than a decade. The basic properties of these effects are
both quantization of the transverse resistance and the nearly complete vanishing of the 
longitudinal resistance. These two properties are observed in more or less all two dimensional
electron gases (2DEG) subjected to an intense magnetic field.
The physical reason for the IQHE is the splitting of the
electronic energy spectrum of a 2DEG into Landau levels in a magnetic field. 
If the Fermi energy lies in the regime of the  localized states
 between two Landau levels one obtains an insulating phase which leads to the IQHE.
In the case of the FQHE one of the Landau levels
is filled by a rational fraction with charge carriers and a gap in the
excitation spectrum is also observed. In an elegant approach the descriptions of the 
IQHE and the FQHE have been unified in the composite fermion picture \cite{3,4}.

The energy gaps which are responsible for the FQHE, are 
the result of electron-electron interaction \cite{5,6}.
Therefore, the properties of these fractional states
depend not only on the Landau level filling factor, but 
also on the interaction strength between the electrons.
Chakraborty and Pietil\"ainen \cite{7} calculated the ground and excited states in the FQHE
and postulated 
that many fractions show a nontrivial behavior. 
For example, their model calculations showed that the ground state of filling factor 
$\rm \frac{2}{3}$ is partly polarized if $\rm B\!<\!7\,T$ but if $\rm B\!>\!15\,T$  
it is fully polarized and in the intermediate regime
the gap vanishes. This behavior has experimentally been observed 
with magneto-resistance measurements in tilted magnetic fields \cite{8,9,10}.
A question that naturally arises is, if one would succeed in modifying the electron-electron
interaction, can other nontrivial effects be found in the FQHE regime?

The interaction strength depends very much on the sample
properties like mobility and finite layer thickness. 
For example,  Zhang and Das Sarma \cite{11} calculated the effects
of a finite layer thickness on the gap in the FQHE states and found 
that the Coulomb interaction is  considerably reduced
if the magnetic length $\rm l_0$ is approximately equal or smaller than the typical thickness of 
the electron wave function in the quantum well. 
According to this theory the fractional energy gap is decreased due to the finite  
thickness of the 2DEG.

Experimental studies of the well-thickness dependence were difficult in the past due to lack of
samples having both high mobility and narrow well thickness. We have succeeded in realizing such
 2DEG structures and have indeed found unexpected
and dramatic resistance structures at several fractional filling factors between 
$\rm \nu = 1$ and $\rm \nu = \frac{1}{2}$. 
At these filling factors the longitudinal magneto-resistance shows narrow 
and large maxima instead of the expected and well established minima.

In this experiment we use a modulation doped  AlGaAs/GaAs structure with a GaAs
quantum  well of $\rm 15\,nm$ thickness; the spacer thickness is $\rm 120\,nm$.
A typical carrier density is about  $\rm 1.3 \!\cdot\! 10^{11} \,cm^{-2} $ after illumination 
 with a LED. At this density the mobility of the sample is $\rm 1.8 \!\cdot\! 10^{6}\, cm^2/Vs$.
The samples are processed in the shape of a standard Hall bar. The contact resistances of the 
In-alloyed ohmic contacts are less than  $\rm 400\,Ohms$ and they 
are not dependent on magnetic field as checked in the IQHE regime. 
Measurements are done on two different Hall bars having
different widths ($\rm 80\, \mu m$ and $\rm 800\, \mu m$).
Four voltage probes along the Hall bar are used to verify the spatial homogeneity 
of the longitudinal resistance. 
The resistance measurements are performed either in a $^3$He bath cryostat at
$\rm 0.4\, K$ or in a dilution refrigerator at $\rm 40\, mK$ using a standard 
ac lock-in technique with a modulation frequency of 23 Hz. 
We also make dc measurements and find identical magneto-resistance traces.
In this letter we show only ac results.

The longitudinal resistance (R$_{xx}$) of the $\rm 80\, \mu m$ Hall bar, measured  
at $\rm 0.4\, K$ with a sweep rate of $\rm 0.7\,T/min$ and a current of $\rm 100 \,nA$,
 is shown by the thin line in Figure 1. 
The resistance shows no unusual behavior at this sweep rate. The minima of the IQHE are 
well developed and in the fractional regime the minimum at $\rm \nu = \frac{2}{3}$ approaches zero.
If however the sweep rate of the magnetic field is reduced to \rm $\rm 0.002\,T/min$, 
a huge longitudinal resistance maximum (HLR) develops very close to the original minimum at 
 $\rm \nu = \frac{2}{3}$. This resistance peak stands out dramatically from the resistance 
values at the surrounding magnetic field regions.
Particularly striking is the sharpness  (width $\rm \Delta B \approx$ $\rm 0.2\,T$) of the HLR.
This anomalous HLR is observed in all studied samples from this wafer.
The position of the maximum remains at filling factor $\rm \nu = \frac{2}{3}$ even if the carrier 
density is varied over the range of  $\rm 1.2\!\cdot\! 10^{11}$ $\rm cm^{-2}$ to $\rm 1.4\!\cdot \!10^{11}$ $\rm cm^{-2}$.

Typical times to form the HLR are determined by setting the appropriate magnetic field 
and recording  R$_{xx}$ as a function of time. Examples are shown in the inset of Figure 1.
It takes $\rm 20\,min$ for the HLR to saturate in case of the  $\rm 80\, \mu m$ Hall bar,
 and several hours for the $\rm 800\, \mu m$ wide one.
These times are longer than the internal electronic relaxation times expected in 
this system.

Figure 2 shows the current dependence of the HLR for two different
samples. The left panel shows the results obtained with  the $\rm 80\, \mu m$ wide Hall bar 
and the right one those obtained with the 
$\rm 800\, \mu m$ wide one. Surprisingly, the height of the HLR depends on the current. 
For the data of the left panel the maximum of the HLR is achieved for current values exceeding 
approximately $\rm 50\,nA$. For the wider samples (right panel) approximately $\rm 400\,nA$  are necessary, corresponding to nearly identical current densities of about $\rm 0.6 \,mA/m$.
In some cases the resistance maximum decreases substantially at higher current densities. The
right panel shows an example.

The fact that the peak resistance of the maxima is usually more than two times larger than the
resistance at the surrounding magnetic field regions rules out any trivial heating effects.
Heating could be responsible for the decrease of the HLR at larger currents
since  the HLR vanishes at bath temperatures exceeding $\rm 0.6\,K$ at all currents. With the same
arguments we can also rule out current-induced breakdown of the QHE.

Unexpected large resistance values in 2DEGs have been reported before in systems which undergo
a metal-insulator phase transition \cite{12}. 
We believe however that this does not occur in our experiment.
First, no resistance maxima nearly as sharp as we observe have been reported.
Second, the current dependence of our maxima does not point to the formation of 
an insulating phase because the resistance should rather increase for smaller currents in case of 
 a metal-insulator transition.
Nevertheless we performed measurements in a dilution refrigerator at $\rm 40\,mK$ to test the 
possibility of an insulating phase. Indeed, the measurements at  $\rm 40\,mK$ do not support
the occurrence of an insulating phase, we rather find a much more complicated behavior of the huge 
longitudinal resistance.


In Figure 3 we show results obtained at $\rm 40\,mK$.
The dashed trace shows the longitudinal resistance as a function  of the  magnetic field
from $\rm 6.5\,T$ (i.e. $\rm \nu = 1$) up to $\rm 12\,T$ at a sweep rate of  
$\rm 0.3\,T/min$. A very regular behavior is observed at
this ''fast'' sweep rate.  The minima at $\rm \nu = \frac{2}{3}$ and 
$\rm \nu = \frac{3}{5}$ are well developed.
The dotted trace shows the results of the same measurement with the magnetic field being swept 
down at the same rate. The curves are markedly different in the magnetic field range from 
$\rm 8.5\,T$ to $\rm 11.5\,T$ (filling factors $\rm \nu = \frac{2}{3}$ 
to $\rm \nu = \frac{1}{2}$). 
The difference is even more pronounced if the down sweep is performed at a slower rate 
such as $\rm 0.006\,T/min$. We observe an anomalous behavior, a huge longitudinal resistance 
 in the down-sweeps at all filling-fractions that are well developed 
between filling factor $\rm \nu = \frac{1}{2}$ and $\rm \nu = 1$ in the fast up-sweep.
We take this again as a signature that the HLR is indeed closely related to the formation 
of the FQHE.
However, it is noteworthy that the FQHE is also well developed at magnetic fields corresponding to 
filling factors below  $\rm \nu = \frac{1}{2}$ and 
to filling factors between $\rm \nu = 1$ and $\rm \nu = 2$, but we cannot find any 
anomalous behavior in those regions so far. 
 The data of Figure 3 show that the qualitative behavior of the HLR is
the same at $\rm 40\,mK$ and at $\rm 0.4\,K$. Particularly, the anomalous resistance value at
 $\rm \nu = \frac{2}{3}$ does not exceed the one at higher temperature, which makes
a metal-insulator transition unlikely. 
Furthermore the dependence of the HLR on current was approximately the same as at higher
temperatures.
In addition the time scale on which the HLR reaches its maximum value is very similar to the
one observed at $\rm 0.4\,K$.

The main difference between the two temperatures is first, that the HLR is observed at more
 fractions and second, that the hysteretic behavior is 
much more pronounced at the lower temperature. For example, when sweeping the magnetic field 
upwards
we usually do not observe the HLR peaks. Exceptions to this are related to the history of 
the sample, for example, if a down sweep was made shortly before.
A slight hysteresis remains at temperatures above about $\rm 250\,mK$:
The width of the HLR peak is smaller when sweeping the magnetic field 
upwards as compared to sweeping downwards.

Measurements of the Hall resistance (R$_{xy}$) reveal that the quantized Hall resistance 
disappears whenever the HLR is observed. There are however deviations from the classical 
(linear) behavior. In the HLR regime the  R$_{xy}$ is approximately 2\% less than the 
classical value when sweeping upwards and approximately 2\% more when sweeping downwards,
i.e. the hysteretic behavior is also shown by the Hall resistance.

It is rather unusual to find such pronounced resistance peaks in magnetic field
regimes where one observes minima and a nearly complete 
disappearance of the longitudinal resistance. The new effect is not just a consequence of the 
very slow sweep rates because we observe the HLR already with standard sweep rates at 
$\rm 40\,mK$.  As far as we can see, the main difference 
between our sample and those which are traditionally used is the reduced thickness of the
quantum well in combination with the high mobility of the 2DEG.  
The importance of the experimental parameters is underlined by our observation that 
the HLR disappears completely if the sample is tilted by $\rm 40^\circ$ against the
 magnetic field direction (Figure 4). Furthermore,  when a carrier density
of $\rm 0.9\!\cdot\! 10^{11}\, cm^{-2}$ was achieved on a different cool-down, the effect likewise 
disappears.

For the HLR to occur, it seems  not to be  enough having a state with vanishing excitation gap.
Because if this were the case we would not expect the  resistance maxima to be larger than the 
resistance in the surrounding magnetic field regions. Actually 
according to \cite{11}, the reduction of the well thickness should lead to an increase of the
excitation gap, i.e. an increased stability of the FQHE, which is contrary to what we observe.
Therefore the explanation of this new effect must be found elsewhere. 

We think it is possible that the electronic system separates spontaneously into different
domains. The high resistance would then be a consequence of the  domain
walls. The formation of domains would rather naturally explain the different time constants 
which we observe in Hall bars of different widths. Possible candidates for such phases are the 
two different ground states for filling factor 
$\rm \nu = \frac{2}{3}$ which were predicted in \cite{7}. 
These two phases differ in total spin of the ground state and in size of the excitation gap.
Let us assume that in our experimental situation the energies of these two ground states are nearly
identical. Then the formation of a domain structure is conceivable.
Our experimental data do indeed strongly support a close connection with the electron spin. First,
we do not observe the HLR at $\rm \nu = \frac{1}{3}$ where theory does not predict the formation of
 competing  ground states. Second, tilting of the sample leads to a rapid disappearance 
of the HLR. This is a strong hint that the HLR is connected with the electron spins
because the additional parallel magnetic field component affects mainly the
Zeeman energy of the electrons.

With such a domain structure, it is necessary
that the spins of some of the electrons flip. Since electron spin flips  are  very 
often connected to nuclear spin flips, it is possible that an electronic
domain structure is related  with a domain structure in the spin configuration of the 
nuclear system. 
Actually our results are very supportive of a close relation between the nuclear spins
with the resistance maxima. Long time constants, of the order of several
minutes to several hours, are very typical of nuclear spins of this type of host lattice 
\cite{13}. 
We assume that the domain structure must be  stabilized by a nuclear spin polarization and 
therefore takes a long time to form. On the other hand an existing nuclear spin polarization
should facilitate the  formation of the electronic domains. This was verified by the 
following experiment:
The magnetic field and the current are set to have the maximal HLR. After the HLR is fully 
developed we sweep the magnetic field fast to a ''waiting'' position where the magnetic field is 
kept constant for times varying from a few seconds to five hours. Then the magnetic field is 
set  back to the original value and   R$_{xx}$  is read immediately. In  this way, one can 
 determine the relaxation time of the HLR as a function of the density of extended electronic
states at the Fermi edge. Results are shown as inset of Figure 3. 
The two sets of data correspond to magnetic fields representing two different longitudinal
 resistances of the sample,
 about $\rm 8.1\,T$ (solid dots) and  $\rm 8.55\,T$ (hollow dots), respectively.
In the first case R$_{xx}$ is  about $\rm 1.5\,k\Omega$ , i.e. the Fermi energy is in the region 
of extended electronic states. One sees that the HLR decays rapidly.
 In contrast, the relaxation time is
of the order of hours if the waiting field is $\rm 8.55\,T$. At this field, the resistance 
is nearly zero and the Fermi energy is in the region of localized states. 
This difference in time constants is exactly what is expected for
the relaxation of nuclear spins via conduction electrons (Korringa effect \cite{14}).

In conclusion we have found sharp resistance maxima at fractional filling factors between 
$\rm \nu = 1 $ and
 $\rm \nu = \frac{1}{2}$ where the resistance tends to vanish in standard samples. 
We suspect that these 
resistance maxima are caused by a domain structure of different electronic spin states connected
with a nuclear spin polarization. This new effect seems to be a consequence of slightly different
experimental parameters, especially the quantum well width, compared to other similar  experiments.

We acknowledge fruitful discussions with R. Gerhardts. M. Riek assisted with the 
preparation of the structures. The help of U. Wilhelm with operating 
the dilution refrigerator system proved to be essential.
This work has been partly supported by the BMBF. 


\begin{figure}
\caption{The longitudinal resistance of a Hall bar at 0.4 K. If the magnetic field
is swept very slowly (0.002T/min, trace "slow") one observes a very
prominent and sharp huge resistance maximum at filling factor 2/3. The
inset shows the temporal evolution of the HLR for two different sample
widths ($\rm 800 \,\mu m $ and $\rm 80 \,\mu m $).\label{Figure1}}
\end{figure}

\begin{figure}
\caption{ Result for Hall-bars of two different width, $\rm 80\, \mu m $ and $\rm 800\, \mu m $, left and right panel respectively. All sweep rates are $\rm 0.002 \,T/min$. The currents are given in the figure. The maximum is most prominent at finite currents which correspond to nearly identical current densities in the two samples. The bold lines correspond to fast sweeps.}
\label{Figure2}
\end{figure}

\begin{figure}
\caption{
Longitudinal resistance measurements at 40 mK ($\rm 80\, \mu m $ width, 100 nA). The HLR
is normally not observed in sweeping the field upwards (dashed trace). It
is already seen in relatively fast down-sweep (0.3 T/min, dotted trace),
but is fully developed at slow down-weeps (0.006 T/min). The HLR can now be
seen also at fractions like 3/5. The inset shows the relaxation of
the HLR via conduction electrons if the sample is temporarily kept at different magnetic fields.}
\label{Figure3}
\end{figure}

\begin{figure}
\caption{R$_{xx}$ for different tilt angles ( $\rm 0^\circ\,,\,23^\circ\,,\,40^\circ$ ); Dashed line: fast sweep, solid line slow sweep. At $\rm 40^\circ$ the effect disappears completely for all currents and for sweep rates down to 0.002 T/min.}
\label{Figure4}
\end{figure}

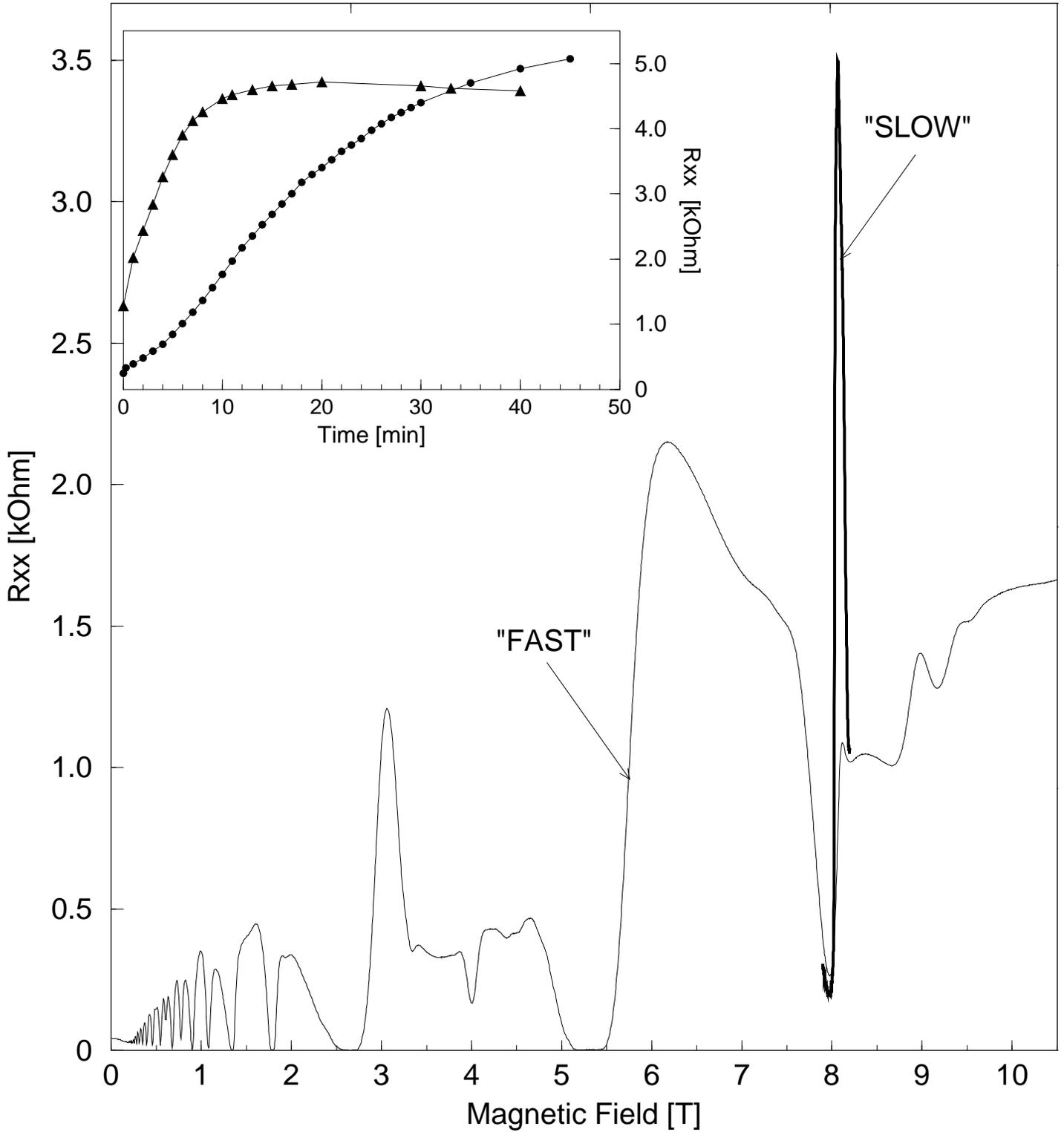
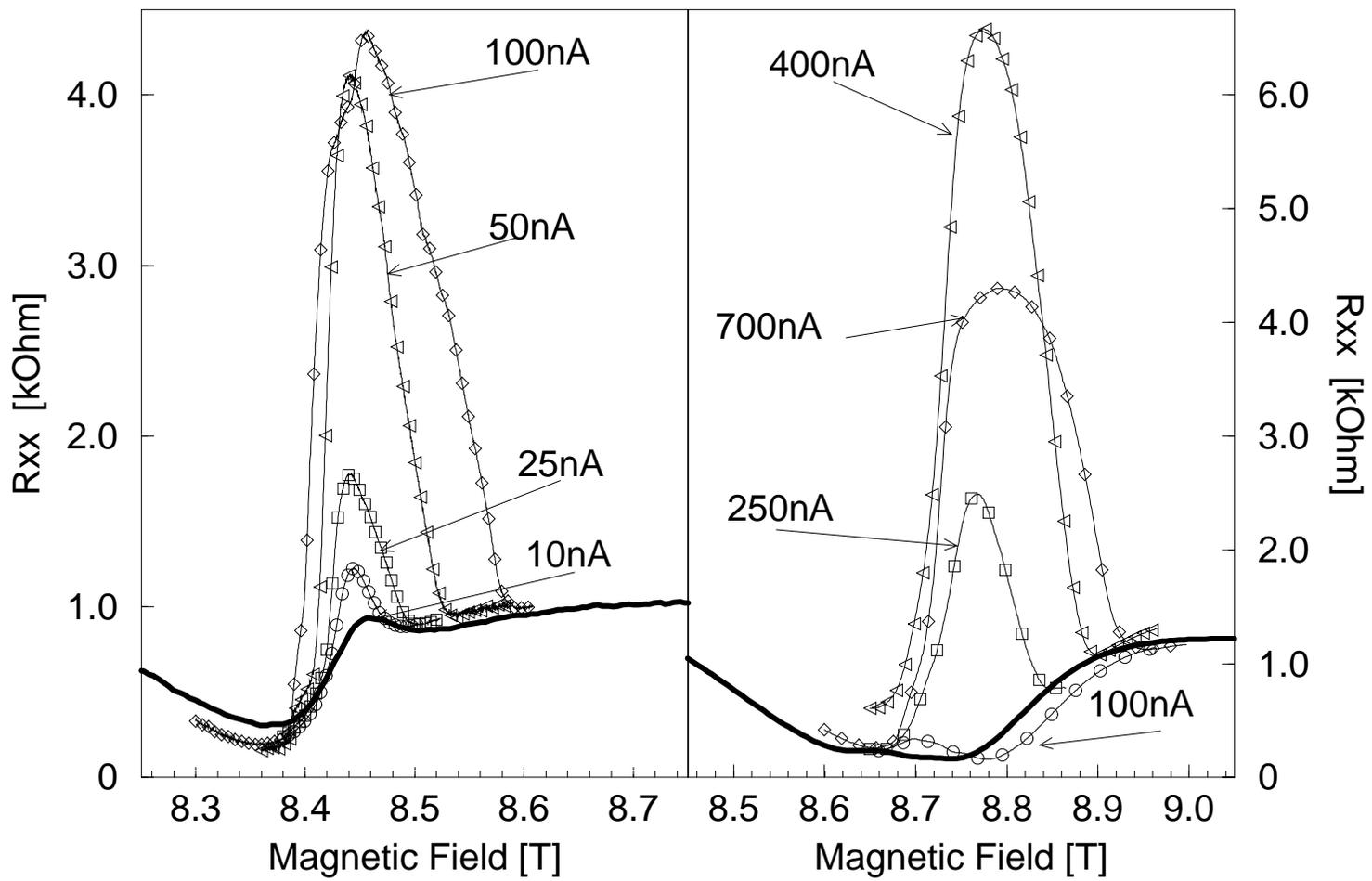
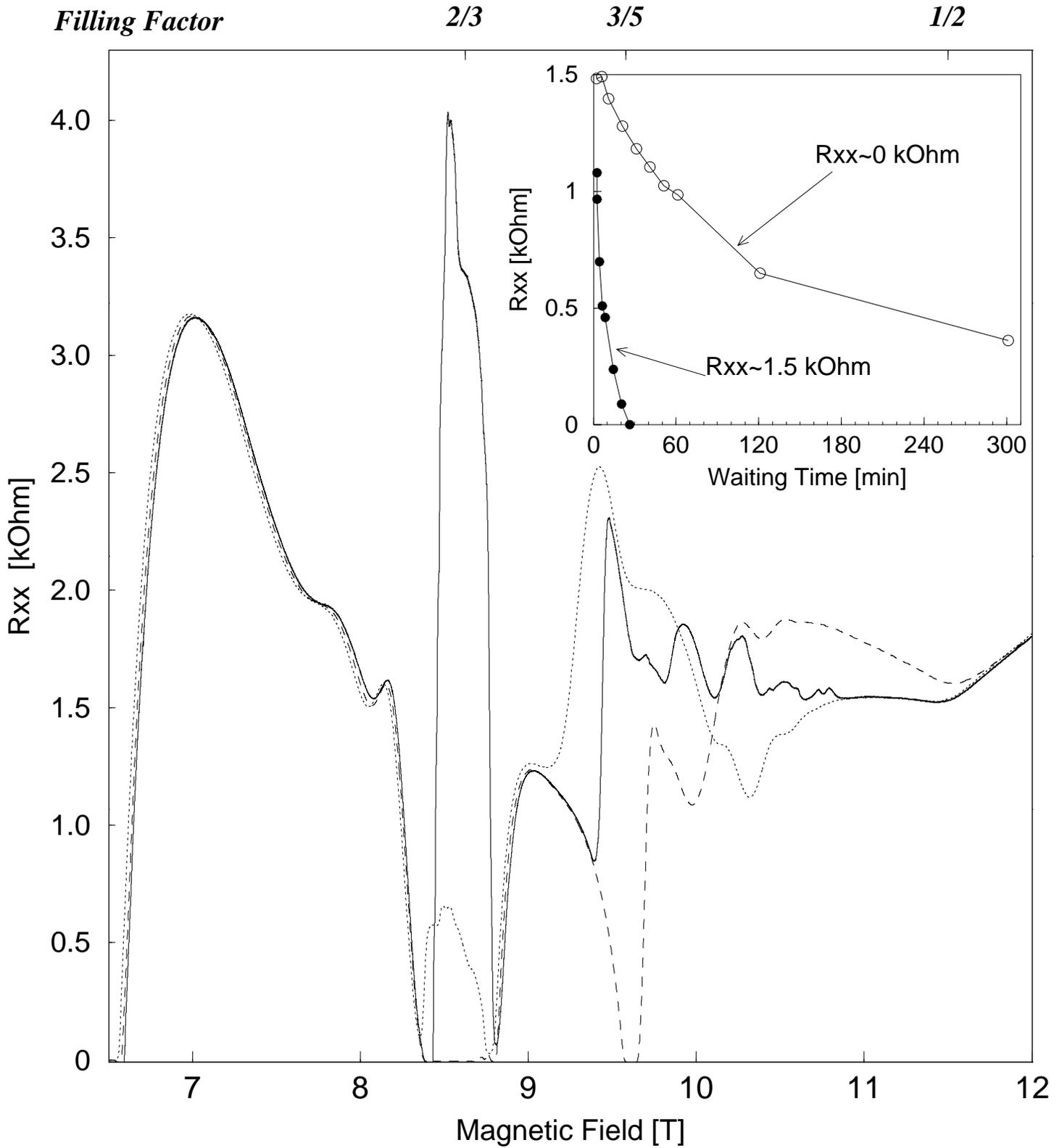
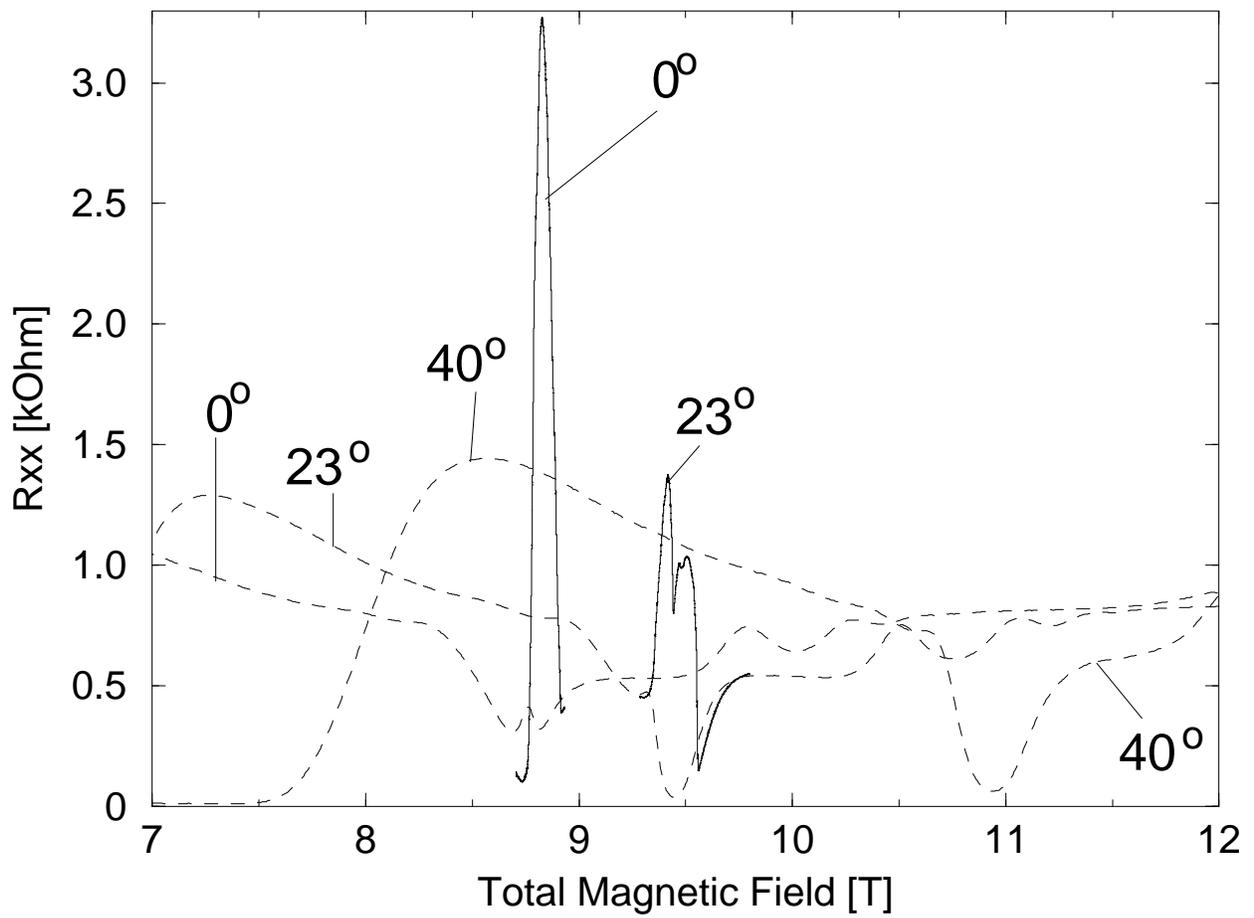
\end{document}